\documentclass{Interspeech2024}

\usepackage{multirow}
\usepackage{subfig}
\usepackage{subcaption}
\usepackage{tablefootnote}
\usepackage[table]{xcolor}
\usepackage{etoolbox}
\usepackage{pgf} 
\definecolor{high}{HTML}{76f013}  
\definecolor{low}{HTML}{ec462e}  
\newcommand*{\opacity}{70}
\newcommand*{\minval}{0.479}
\newcommand*{\maxval}{0.679}
\newcommand{\g}[1]{
    \ifdimcomp{#1pt}{>}{\maxval pt}{#1}{
        \ifdimcomp{#1pt}{<}{\minval pt}{#1}{
            \pgfmathparse{int(round(100*(#1/(\maxval-\minval))-(\minval*(100/(\maxval-\minval)))))}
            \xdef\tempa{\pgfmathresult}
            \cellcolor{high!\tempa!low!\opacity} #1
    }}
}

\interspeechcameraready

\title{
Simulating articulatory trajectories with phonological feature interpolation
}

\name[affiliation={1,2}]{Angelo}{Ortiz Tandazo}
\name[affiliation={3}]{Thomas}{Schatz}
\name[affiliation={2}]{Thomas}{Hueber}
\name[affiliation={1, 4}]{Emmanuel}{Dupoux}

\address{
  $^1$ENS, PSL Research University, EHESS, CNRS, France\\
  $^2$Univ. Grenoble Alpes, CNRS, Grenoble INP, GIPSA-lab, France \\
  $^3$Aix-Marseille Univ., CNRS, LIS, France \\
  $^4$Meta AI
}
\email{
angelo.ortiz.tandazo@ens.psl.eu,
thomas.schatz@univ-amu.fr,
thomas.hueber@grenoble-inp.fr,
emmanuel.dupoux@gmail.com
}


\begin{document}

\maketitle


\begin{abstract}

    As a first step towards a complete computational model of speech learning involving perception-production loops, we investigate the forward mapping between pseudo-motor commands and articulatory trajectories. Two phonological feature sets, based respectively on generative and articulatory phonology, are used to encode a phonetic target sequence. Different interpolation techniques are compared to generate smooth trajectories in these feature spaces, with a potential optimisation of the target value and timing to capture co-articulation effects.
    We report the Pearson correlation between a linear projection of the generated trajectories and articulatory data derived from a multi-speaker dataset of electromagnetic articulography (EMA) recordings. A correlation of 0.67 is obtained with an extended feature set based on generative phonology and a linear interpolation technique. We discuss the implications of our results for our understanding of the dynamics of biological motion. 
    
\end{abstract}

\section{Introduction}

Recent advances in self-supervised learning (SSL) have led to progress in various speech processing tasks \cite{yang21c_interspeech, dunbar2022self} and language modelling from speech units \cite{borsos2023audiolm}.
These SSL models require increasingly greater amounts of (unlabelled) data to capture as much acoustic variance as possible.
Moreover, their capacity to learn high-level language representations hinges on the quality of the underlying speech units \cite{lakhotia2021generative}, which are not linguistically interpretable \cite{dunbar2022self,lavechin2023acquisition}.
Importantly, these representations remain sensitive to contextual effects such as co-articulation \cite{hallap2023evaluating} making them sub-optimal to efficiently code context-invariant phonological units.

According to the motor \cite{liberman1985motor} or perceptuo-motor theories \cite{schwartz2012perception} of speech perception, humans' `quest for invariance' \cite{perkell1986invariance} is done by recovering motor or articulatory representations from an auditory input.
These representations are supposed to be less variable than their acoustic counterparts. Several studies have found neuro-physiological correlates of this mental `sensory-to-motor' inverse mapping in speech perception \cite{SKIPPER201777}.     
Incorporating such motor representation into SSL speech models could potentially improve their performance (\textit{e.g.} noise robustness, low-resource downstream tasks, etc.) and lead to more plausible computational models of speech and language acquisition.

\begin{figure}[t]
  \centering
  \includegraphics[width=\linewidth]{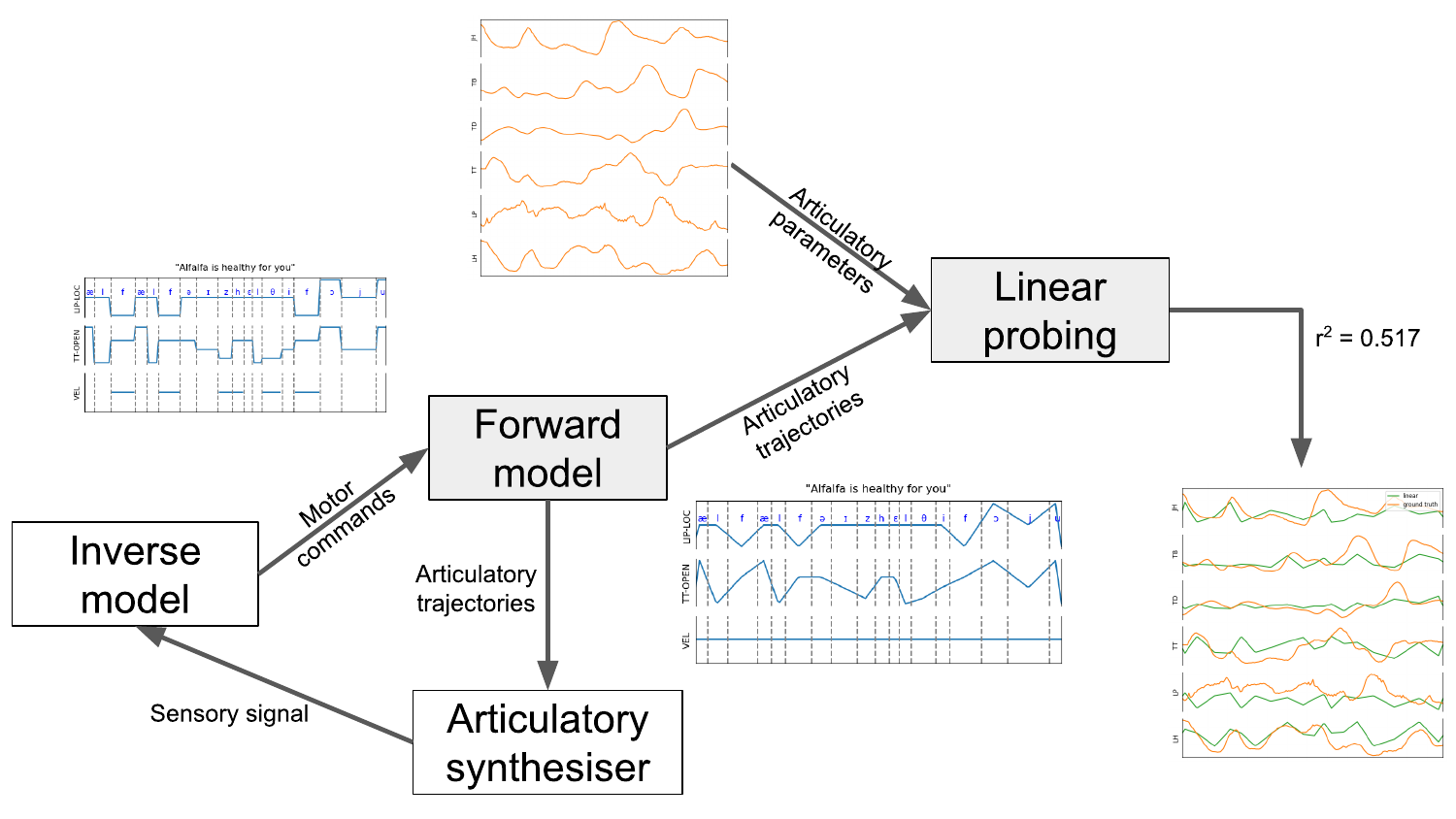}
  \caption{Simplified diagram of a speech perception-production loop (to the left). The focus of this work lies in the forward model and the linear probing (to the right).}
  \label{fig:speech_model}
\end{figure}

In a simplified perception-production loop of speech motor control \cite{jordan1999computational} (see the bottom left of Figure \ref{fig:speech_model}), the motor commands are derived from the sensory, acoustic signal and used to generate the underlying articulatory trajectories (via a so-called \textit{forward model}).

Several computational models of such speech perception-production loop have been proposed in the literature \cite{patri2015optimal, philippsen2014learning, rasilo2017online, georges_et_al_icassp2022}. However, in most studies, the motor or articulatory representations are derived from a specific speaker or a specific articulatory model.
As a result, these models, while enabling to study some of the underlying processes of speech acquisition, perception and motor control by simulation, are not designed to scale up to large numbers of speakers or languages.

As a first step towards a more universal SSL speech model integrating motor or articulatory representations, we focus in the present study on the forward model, \textit{i.e.} from the motor commands to the generation of articulatory trajectories.
First, we investigate different feature sets to encode a given phonetic target sequence, by relying either on the generative phonology (GP, \cite{chomsky1968sound}) or on the articulatory phonology (AP, \cite{browman1992articulatory}) theories.
Importantly, a phonetic target is here encoded in terms of phonologically-motivated and articulatory-related categories (\textit{e.g.} the place of articulation for a consonant in GP, the location and degree of a constriction in AP). 

To generate continuous (and smooth) trajectories in these feature spaces (with a forward model), we test different interpolation methods, which differ from the dynamic properties desired at each phonetic target (\textit{e.g.} zero and/or continuous velocity at each target). 
To account for the uncertainty of our timing heuristic and the (potential) target undershoot phenomenon, we also consider variants performing target optimisation both in space (find the offset from the ideal feature value, \textit{e.g.} lips only partly closed) and time (reach a target sooner or later).
Moreover, the proposed approach can deal with unspecified features for which the value depends on the context (for instance, the position of a constriction modulated by the vocalic context).
The generated trajectories in the GP or AP feature spaces are evaluated using a linear probing technique.
A linear model is learnt between the generated trajectories on the one hand, and the parameters of an articulatory model \cite{maeda2012art} built from a multi-speaker electromagnetic articulography (EMA) dataset on the other.
Such an evaluation was also used in \cite{cho2023probing} to probe HuBERT representations, among other SSL models.  

The main contributions and findings of the paper are the following:
\begin{enumerate*}[label=(\roman*)]
    \item we propose a general methodology to probe pseudo-motor commands and forward models in a computational model of a speech perception-production loop;
    \item we show that features derived from generative phonology (GP) correlate better with real articulatory recordings than articulatory phonology (AP) ones;
    \item a bit surprisingly, we show that a linear interpolation between these features better captures the dynamics of real articulatory data compared to a more complex one (spline based), with constraints on the velocity and/or acceleration at each phonetic target;
    \item we show that the use of unspecified (context-dependent) phonological features improves performance, probably by allowing the forward model to better account for natural co-articulation patterns. 
\end{enumerate*}


The code with the data processing and interpolation methods can be found at
\ifinterspeechfinal
\url{https://github.com/angelo-ortiz/articulatory-probing}.
\else
[to ensure author anonymity, the link to the resource will be added after the review process].
\fi

\section{Methodology}

\subsection{Phonological feature set}
\label{section:phono_feature}

Two phonological feature sets were used.
The first one, based on generative phonology and referred to as the GP feature set, was proposed by \cite[Chapter 4]{hayes2008introductory}.
It describes phonetic targets in terms of $ 26 $ manner, laryngeal and place features. 
The GP feature set is considered under two variants: with unknown support and binary.
Following \cite[p.\ 91]{hayes2008introductory}, some phonemes have zero-value features, notably because their values depend on their local context within an utterance or simply because they are irrelevant to the underlying phoneme.
Hence, the ternary-valued (including the zero values) GP feature set is considered as is (to be used by interpolating methods handling unknown values) but also in a binary form, in which a zero value is considered as being the `absence' of the given feature (thus, negatively valued).

The second feature set is based on articulatory phonology (AP), and the location and degrees of constriction of 5 major articulators in the vocal tract.
AP-based features have been successfully used in automatic speech recognition, first within a Bayesian framework  \cite{livescu2005feature} to deal with pronunciation variability in spontaneous speech, and then in a DNN-based system \cite{BADINO2016173} to increase the robustness to noise.
The AP feature values come in the form of categorical distributions over totally ordered categories \cite[p.\ 126]{livescu2005feature}.
Feature values are typically Dirac distributions, except for some phoneme features that depend on the phonemic context.
Similarly to the GP feature set, we have an AP unknown variant by considering the non-Dirac distributed feature values as \textit{unknown} values to be found contextually.
This feature set is then used in a scalar version, in which each feature-value category is mapped to a real value (scalar AP: 8 features); and a one-hot version, in which the feature-value categories are one-hot encoded (one-hot AP: 32 features).

\begin{table}[t!]
  \caption{Articulatory score: average Pearson correlation coefficients of the $ 6 $ articulatory parameters and $ 6 $ speakers. The scores correspond to each interpolation method's best configuration. Default: binary features without optimisation. Variants: unknown features$^\mu$, timing optimisation$ ^\dagger $, timing and position optimisation$ ^\ddagger $. (Standard error across the 6 different speakers was found to be $0.01$ on average.)
  }
  \label{tab:scores-interp}
  \centering
  \begin{tabular}{ c c l r }
    \toprule
    Feature set & \# Features & Method & \multicolumn{1}{c}{Score $ \uparrow $} \\
    \midrule
    \midrule
    \multirow{4}{2cm}{\centering GP + one-hot phoneme} & \multirow{3}{*}{$ 73 $} & piecewise-cst & $\g{0.595}$~~~ \\
    & & linear$^\mu$ & $\mathbf{\g{0.679}}$~~~ \\
    & & cubic Hermite$^\mu$ & $\g{0.668}$~~~ \\
    & & natural cubic$^\ddagger$ & $\g{0.663}$~~~ \\
    \midrule
    \multirow{3}{2cm}{\centering one-hot AP + one-hot phoneme} & \multirow{3}{*}{94} & linear$^\mu$ & $\g{0.663}$~~~ \\
    & & cubic Hermite$^\mu$ & $\g{0.648}$~~~ \\
    & & natural cubic$^\mu$ & $\g{0.628}$~~~ \\
    \midrule
    \multirow{3}{2cm}{\centering scalar AP + one-hot phoneme} & \multirow{3}{*}{70} & linear$^\mu$ & $\g{0.656}$~~~ \\
    & & cubic Hermite$^\mu$ & $\g{0.642}$~~~ \\
    & & natural cubic$^\mu$ & $\g{0.624}$~~~ \\
    \midrule
    \multirow{4}{2cm}{\centering one-hot phoneme} & \multirow{3}{*}{$ 47 $\tablefootnote{British long vowels and the silence are included.}} & piecewise-cst & $\g{0.589}$~~~ \\
    & & linear & $\g{0.645}$~~~ \\
    & & cubic Hermite$^\ddagger$ & $\g{0.642}$~~~ \\
    & & natural cubic$^\ddagger$ & $\g{0.630}$~~~ \\
    \midrule
    \multirow{4}{*}{GP} & \multirow{4}{*}{$ 26 $} & piecewise-cst & $\g{0.559}$~~~ \\
    & & linear$^\mu$ & $\g{0.630}$~~~ \\
    & & cubic Hermite$^{\mu(\dagger)}$ & $\g{0.622}$~~~ \\
    & & natural cubic$^\ddagger$ & $\g{0.629}$~~~ \\
    \midrule
    \multirow{3}{*}{one-hot AP} & \multirow{3}{*}{$ 32 $} & linear$^\mu$ & $\g{0.608}$~~~ \\
    & & cubic Hermite$^{\mu\ddagger}$ & $\g{0.596}$~~~ \\
    & & natural cubic$^\mu$ & $\g{0.538}$~~~ \\
    \midrule
    \multirow{3}{*}{scalar AP} & \multirow{3}{*}{$ 8 $} & linear$^\mu$ & $\g{0.511}$~~~ \\
    & & cubic Hermite$^{\mu\ddagger}$ & $\g{0.506}$~~~ \\
    & & natural cubic$^\mu$ & $\g{0.479}$~~~ \\
    \bottomrule
  \end{tabular}
\end{table}

To ensure that the feature-level information for phonemes is relevant, we also use a \textit{feature} set with one-hot phoneme encodings. In total, we evaluate seven feature sets: one-hot phonemes, scalar AP (also enriched with one-hot phonemes), one-hot AP (also enriched with one-hot phonemes), binary GP and unknown-supporting GP (also enriched with one-hot phonemes).

\subsection{Dataset}

The articulatory data comes from the publicly available MOCHA-TIMIT dataset\footnote{The dataset can be found at \url{https://www.cstr.ed.ac.uk/research/projects/artic/mocha.html}}.
It provides electromagnetic articulography (EMA) recordings for 460 short sentences read by 8 British English speakers along 12 dimensions (2D midsagittal coordinates for 6 articulators: tongue tip, tongue body, tongue dorsum, lower incisor, upper lip and lower lip.)
In this study, we consider 6 speakers, namely \texttt{fsew0}, \texttt{msak0}, \texttt{ffes0}, \texttt{mjjn0}, \texttt{faet0} and \texttt{maps0}, because a sequence of waveform files did not match the given transcriptions for the other two speakers.
For each speaker, the 460 utterances are split into 410 for training (out of which 20 are randomly drawn for development) and 50 for testing.

The EMA data is low-pass filtered at \SI{50}{\hertz} and down-sampled from \SI{500}{\hertz} to \SI{100}{\hertz}.
As with \cite{serrurier2012tongue}, raw EMA data is then converted into an easier-to-interpret and lower-dimensional set of 6 `articulatory parameters' (jaw height, tongue body, tongue back, tongue tip, lip protrusion and lip height) using a linear decomposition technique, which is often referred to as `guided PCA'. It aims to decouple the jaw from tongue and lip movements and extract independent degrees of freedom from the vocal tract.

Each audio recording was segmented at the phonetic level using the Montreal Forced Aligner\footnote{https://github.com/MontrealCorpusTools/Montreal-Forced-Aligner}.
Based on the resulting phonetic segmentation, the initial and final utterance silences were removed and the utterances with non-silence boundary phones were discarded.
The filtered phonetic segmentations were later mapped into \textit{featural} segmentations by replacing the phones with the phonological features of their underlying phonemes (lookup table mapping).
Finally, we inferred timings for the phonological targets from the time midpoint of each phoneme in the featural segmentation.

\subsection{Forward model}
\label{section:forward_model}

Different forward mapping techniques are tested to generate continuous trajectories from the discrete pseudo-motor commands provided by GP and AP (Section \ref{section:phono_feature}).
As a first baseline, we use the piecewise-constant interpolation, which keeps all the phonological features constant for the duration given by the phonetic segmentation.

To test a smoothness degree that better fits the articulatory space, we test linear and cubic interpolation methods.
Specifically, we consider two cubic methods: the cubic Hermite spline and the natural cubic spline.
The former enforces zero velocity at all targets, and continuity of both position and velocity (so the acceleration is possibly discontinuous at the targets); whereas the latter enforces continuity of all position, velocity and acceleration, with zero acceleration at the initial and final targets.

Formally, let $ d \in \mathbf{N}_{>0} $ be the number of phonological features.
For a given utterance, let $ K \in \mathbf{N}_{>0} $ be its number of non-boundary (or intermediate) targets.
Then, its featural segmentation is denoted by $ (\mathbf{X}, \mathbf{Y}) \in \mathbf{R}^{(K+2) \times d} \times \mathbf{R}^{(K+2) \times 2} $, where $ \mathbf{X} $ contains the $ K+2 $ target positions, and $ \mathbf{Y} $ the targets' time intervals.
In this work, we remove the boundary silences, so $ y_{1, \ast} = \mathbf{0}_2 $, $ y_{K+2, \ast} = y_{K+1, 2} \mathbf{1}_2  $, and $ x_{1, \ast} = x_{K+2, \ast} = \mathbf{0}_d $.
From this, we deduce a vector of midpoint target timings $ \mathbf{t} \triangleq \frac{1}{2} Y \mathbf{1}_2  $.
The (base) interpolating function for the utterance $ (\mathbf{X}, \mathbf{Y}) $, expressed as $ f(\tau; \mathbf{X}, \mathbf{t}) $, $ 0 \leq \tau \leq t_{K+2} $, thus satisfies
\begin{equation}
    f(t_k; \mathbf{X}, \mathbf{t}) = x_{k, \ast},
    \label{eq:interp}
\end{equation}
for all $ 1 \leq k \leq K+2 $.

To tackle the uncertainty of the midpoint-timing heuristic assumed for the base interpolating functions and to allow for target undershoot\footnote{Target undershoot occurs when there is not enough time for the forward model to reach some targets.}, we include two additive optimisations over time and space. The optimised interpolating function $ f $ learns the target positions and timings $ (\mathbf{X}', \mathbf{t}') $ such that
\begin{equation}
    f(\tau; \mathbf{X}, \mathbf{t}) = g(\tau, \mathbf{X}', \mathbf{t}'), 0 \leq \tau \leq t_{K+2},
    \label{eq:optim-interp}
\end{equation}
where the base interpolating function $ g $ satisfies Equation \ref{eq:interp}, by minimising
the objective function 
\begin{align}    
    L_{\lambda}(\mathbf{X}', \mathbf{t}') \triangleq & \int_{0}^{t_{K+2}} \lVert g''(\tau; \mathbf{X}', \mathbf{t}') \rVert_2^2 \,d \tau \nonumber \\
    &+ \lambda \sum_{k=2}^{K+1} \lVert g(t_k'; \mathbf{X}', \mathbf{t}') - x_{k,\ast} \rVert_2^2. 
    \label{eq:target-optim}
\end{align}
We optimise the objective function $ L_{\lambda} $ by on-utterance gradient descent with the initialisation $ (\mathbf{X}', \mathbf{t}') = (\mathbf{X}, \mathbf{t}) $.

\subsection{Linear probing}

To evaluate the different interpolation methods described above, we adopted the same metric as used in \cite{cho2023probing}, referred to as the \textit{articulatory score}.

Let $ S_i = \{ (\mathbf{X}_j, \mathbf{t}_j, \mathbf{Z}_j)_j \} $ be the set of utterances for the $ i $th speaker, with $ \mathbf{Z}_j \in \mathbf{R}^{n_{\mathbf{Z}_j} \times 6} $ being the $ n_{\mathbf{Z}_j} $ articulatory parameters of the $ j $th utterance.
Then, for each interpolation method $ f $ and speaker $ S_i $, we learn a linear transformation $ h_i $ that minimises the reconstruction loss from the interpolated articulatory trajectories and the expected articulatory parameters as follows
\begin{equation}
    \mathcal{L}_i = \frac{1}{\lvert S_i \rvert}\sum_{(\mathbf{X}, \mathbf{t}, \mathbf{Z}) \in S_i} \frac{1}{n_{\mathbf{Z}}} \sum_{k=1}^{n{_\mathbf{Z}}} \left\lVert h_i\left(f\left(\frac{k}{100}; \mathbf{X}, \mathbf{t}\right)\right) - z_{k, \ast} \right\rVert_2^2.
    \label{eq:linear-probe}
\end{equation}

This is done via gradient descent with a learning rate of $ \num{0.001} $ via the Adam optimiser, with $ \beta = (0.9, 0.999) $.
We run the learning procedure for $ 100 $ epochs unless it is early stopped when the validation loss stops decreasing (patience fixed at $ 5 $).

For the optimised cubic interpolations, we first do a grid search of the hyper-parameters used in the on-utterance target optimisation:
\begin{enumerate*}[label=(\roman*)]
    \item timing learning rate within $ \{ \num{1e-6}, \num{5e-6}, \num{1e-5}, \num{5e-5}, \num{1e-4} \} $,
    \item position learning rate within $ \{ \num{1e-3}, \num{1e-2}, \num{1e-1} \} $, and
    \item loss weight parameter $ \lambda \in \{0,  \num{1e3}, \num{1e4}, \num{1e5}, \num{1e6}, \num{1e7} \} $\footnote{The spatial term of the loss in Equation \ref{eq:target-optim} is very small compared with the smoothness term.}.
\end{enumerate*}
For each interpolation method, the hyper-parameter configuration with the highest articulatory score in the development set was selected. 
Finally, we compute the Pearson correlation coefficient (PCC) between the learnt linear projections and the articulatory parameters. The final articulatory score of an interpolation method is then the average PCC of all articulatory parameters and speakers.

\section{Results}

\renewcommand*{\minval}{0.53}
\renewcommand*{\maxval}{0.75}

We run the linear, cubic Hermite spline and natural cubic spline interpolation methods (Section \ref{section:forward_model}) on the seven feature sets derived from one-hot phoneme encoding, AP and GP theories (Section \ref{section:phono_feature}).
The piecewise-constant interpolation can only be run on fully specified feature sets, here the one-hot phoneme and binary GP feature sets.

\begin{table*}[t!]
    \centering
    \caption{Articulatory score for each speaker on the GP feature set enriched with one-hot phonemes.}
    \label{tab:score-speaker}
    \begin{tabular}{ c *{6}{c} c }
        \toprule
        Method & \texttt{msak0} & \texttt{fsew0} & \texttt{ffes0} & \texttt{mjjn0} & \texttt{maps0} & \texttt{faet0} & Average \\
        \midrule \midrule
        piecewise-cst & $\g{0.634} $ & $\g{0.616} $ & $\g{0.590} $ & $\g{0.591} $ & $\g{0.537}$ & $\g{0.604}$ & $\g{0.595}$ \\
        linear & $ \mathbf{\g{0.729}} $ & $ \mathbf{\g{0.704}} $ & $ \mathbf{\g{0.666}} $ & $ \mathbf{\g{0.658}} $ & $ \mathbf{\g{0.623}} $ & $ \mathbf{\g{0.693}} $ & $ \mathbf{\g{0.679}} $ \\
        cubic Hermite & $\g{0.718} $ & $\g{0.695} $ & $\g{0.655} $ & $\g{0.649} $ & $\g{0.611} $ & $\g{0.681}$ & $\g{0.668}$ \\
        natural cubic & $\g{0.711} $ & $\g{0.686} $ & $\g{0.652} $ & $\g{0.632} $ & $\g{0.613} $ & $\g{0.685} $ & $\g{0.663} $ \\
        \bottomrule
    \end{tabular}
    
    \bigskip
    
    \caption{Articulatory score for each articulatory parameter on the GP feature set enriched with one-hot phonemes.}\label{tab:score-articulator}
    \begin{tabular}{ c *{6}{c} c }
        \toprule
        \multirow{2}{*}{Method} & \multirow{2}{1cm}{\centering Jaw height} & \multirow{2}{1cm}{\centering Tongue body} & \multirow{2}{1cm}{\centering Tongue dorsum} & \multirow{2}{1cm}{\centering Tongue tip} & \multirow{2}{1cm}{\centering Lip protrusion} & \multirow{2}{1cm}{\centering Lip height} & \multirow{2}{1cm}{\centering Average} \\ \\
        \midrule \midrule
        piecewise-cst & $\g{0.646} $ & $\g{0.653} $ & $\g{0.543} $ & $\g{0.539} $ & $\g{0.532} $ & $\g{0.658} $ & $\g{0.595} $ \\
        linear & $ \mathbf{\g{0.715}} $ & $ \mathbf{\g{0.750}} $ & $ \mathbf{\g{0.625}} $ & $ \mathbf{\g{0.627}} $ & $ \mathbf{\g{0.621}} $ & $ \mathbf{\g{0.736}} $ & $ \mathbf{\g{0.679}} $ \\
        cubic Hermite & $\g{0.703} $ & $\g{0.742} $ & $\g{0.618} $ & $\g{0.616} $ & $\g{0.610} $ & $\g{0.722} $ & $\g{0.668} $ \\
        natural cubic & $\g{0.708} $ & $\g{0.733} $ & $\g{0.604} $ & $\g{0.606} $ & $\g{0.615} $ & $\g{0.713} $ & $\g{0.663} $ \\
        \bottomrule
    \end{tabular}
\end{table*}

Table \ref{tab:scores-interp} reports the articulatory score by feature space and interpolation method on a held-out test.
The scores correspond to each interpolation method's best configuration on each feature set.
We observe that the scalar AP proves to be (very) difficult to interpolate on, but replacing the fixed values (probably needing learning) with equidistant values in the form of one-hot encodings helps close the gap to the GP features.
Surprisingly, the one-hot phoneme encodings are the best \textit{single} feature set.

Given the potential complementarity of information between the GP/AP feature sets and the one-hot phoneme encodings, we probe the GP and AP features enriched with the latter.
The mix turns out to be beneficial for all the interpolation methods, although we lose the reduced number of interpretable features sought for the inverse models in perspective. 

Tables \ref{tab:score-speaker} and \ref{tab:score-articulator} show the articulatory scores per speaker and articulatory parameter, respectively, on the best feature space, namely GP features enriched with one-hot phonemes.
In both cases, the ranking induced by the average score (linear $ \succ $ cubic Hermite $ \succ $ natural cubic $ \succ $ piecewise constant) is met throughout the conditions, bar the two speakers \texttt{maps0} and \texttt{faet0}, the jaw height and the lip protrusion (natural cubic $ \succ $ cubic Hermite).
Interestingly, from Tables \ref{tab:scores-interp}, \ref{tab:score-speaker} and \ref{tab:score-articulator}, it is clear that the linear interpolation method better exploits the given phonological spaces, regardless of the feature nature, speaker or articulatory parameter.

In Table \ref{tab:scores-interp}, we see that most of the best scores reported were obtained on features with unknown support.
The results in Table \ref{tab:scores-learning} support the hypothesis that, in general, keeping and interpolating unknown feature values is better than associating them with a fixed value.
On the other hand, the effect of the target timing and/or position optimisation depends on the interpolation method.
For instance, when we optimise the timings on the cubic Hermite spline, the articulatory score does not improve, and the spatial optimisation has the same (negative) impact. This is why we see few cubic Hermite interpolations with target optimisations in Table \ref{tab:scores-interp}.

\begin{table}[th]
  \caption{Comparison of GP + one-hot phoneme feature set variants and the effect of target optimisation. The two leftmost scores correspond to the binary and unknown-supporting feature-set variants without optimisation, and the right scores to the timing-only and the timing-and-position optimisations on the underlined feature sets.}
  \label{tab:scores-learning}
  \centering
  \begin{tabular}{ c c c c c }
    \toprule
    \multirow{2}{*}{Method} & \multicolumn{2}{c}{Non-optimised} & \multicolumn{2}{c}{Optimised} \\
    & \multicolumn{1}{c}{Binary} & \multicolumn{1}{c}{Unknown} & \multicolumn{1}{c}{Time} & \multicolumn{1}{c}{Time \& space} \\
    \midrule \midrule
    linear & $0.659$ & $\mathbf{0.679}$ & & \\
    cubic Hermite & $0.645$ & $\underline{\mathbf{0.668}}$ & $0.660$ & $0.649$ \\
    natural cubic & $\underline{0.623}$ & $0.638$ & $0.624$ & $\mathbf{0.663}$ \\
    \bottomrule
  \end{tabular}
\end{table}

\section{Conclusion}

In this study, we have analysed phonological features as potential pseudo-motor commands in a computational model of a speech perception-production loop. We found that:
\begin{enumerate*}[label=(\roman*)]
    \item smooth trajectories on generative phonology features correlate better with articulatory parameters than those on articulatory phonology ones, with a correlation coefficient of $ 0.67 $ when GP features are enriched with one-hot phoneme encodings,
    \item a linear forward model better captures the dynamics of real articulatory data, but target optimisation (in terms of timing and/or position) helps a cubic model to reduce the gap,
    \item with the AP features, a better correlation coefficient is obtained with a one-hot encoding, in which all the values for a given feature are equidistant to one another, rather than with a fixed, scalar continuous one,
    \item interpolating unknown (or context-dependent) features is better than associating a fixed value with them.
\end{enumerate*}

Since interpolating under-specified dimensions of articulatory targets appears to lead to a better fit, future work could try to push this strategy further by incorporating more under-specified dimensions in featural segmentations, thus enabling the smoothness of the forward model's trajectories to better model co-articulation.

Further work should also investigate the reason why linear interpolation of articulatory targets outperforms smoother cubic spline interpolation.
This is surprising since articulatory trajectories are not linear, cubic splines are excellent interpolators and the biological motion literature suggests that smooth trajectories should fit articulatory data well \cite{flash1985coordination}.
A possible interpretation is that unwarranted assumptions in our analyses cause the observed advantage of linear interpolation methods.
Based on our results, the advantage of linear interpolation does not appear sensitive to assumptions regarding target definition (generative vs articulatory, specification, timing, position).
It may be the case, however, that the assumption of a fixed inventory of targets at the phonemic level is too optimistic, even for a single speaker, at least without controlling further variables that may modulate target parameters, such as prosodic effects.
To test this hypothesis, simulated trajectories could be used to determine if ignoring prosodic effects would predict an advantage for linear interpolation even when using a cubic spline model for the dynamics.
Alternatively, our results may indicate that classical results on biological motion, obtained in highly controlled settings, do not accurately characterize the dynamics of biological motion in less restricted environments.
To test this hypothesis, our methodology could be applied to more controlled trajectories (e.g. isolated syllables), where it should find that smoother trajectories provide a better fit than linear interpolation.
This would validate our methodology, which could then be leveraged to better understand the dynamics of biological motion in the wild.

\ifinterspeechfinal

\section{Acknowledgements}
This work was granted access to the HPC resources of IDRIS under the allocation 2023-AD011014739 made by GENCI.
\else
\fi

\bibliographystyle{IEEEtran}
\bibliography{mybib}

\end{document}